\documentclass{PoS}
\title{Vacuum Expectation Values of Twisted Mass Fermion Operators\thanks{This work was partially supported by the National Science Foundation, Computational Mathematics Program, under grant 0310573, by the Natural Sciences and Engineering
Research Council of Canada, and by the Canada Research Chairs Program.}}

\ShortTitle{Vacuum Expectation Values of Twisted Mass Fermion Operators}

\author{Abdou Abdel-Rehim\\
        Department of Physics, Baylor University, Waco, TX 76798-7316, USA.\\
        E-mail: \email{Abdou$\_$Abdel-Rehim@baylor.edu}}

\author{Randy Lewis\\
        Department of Physics, University of Regina, Regina, SK S4S 0A2, Canada.\\
        E-mail: \email{randy.lewis@uregina.ca}}
        
\author{\speaker{Walter Wilcox}\\
        Department of Physics, Baylor University, Waco, TX 76798-7316, USA.\\
        E-mail: \email{Walter$\_$Wilcox@baylor.edu}}

\abstract{
Using noise methods on a quenched $20^3 \times 32$ lattice at $\beta=6.0$, we have investigated vacuum expectation values and relative linear correlations among the various Wilson and twisted mass scalar and pseudoscalar disconnected loop operators. We show results near the maximal twist lines in $\kappa$, $\mu$ parameter space, either defined as the absence of parity mixing or the vanishing of the PCAC quark mass. }

\FullConference{The XXV International Symposium on Lattice Field Theory\\
                 July 30 - August 4 2007\\
                 Regensburg, Germany}
\begin{document}

\section{Introduction}
Twisted mass (TM) fermions have great potential in extending lattice
QCD calculations to smaller quark masses. However, their
disconnected loop properties are not well explored. We have done a
number of calculations of quenched TM quark-loop properties, using real $Z(2)$ noises and
the efficient GMRES-DR/GMRES-Proj inversion algorithms~\cite{PoS1,PoS2}, concentrating on the scalar-pseudoscalar sector, to determine their mixing patterns and other properties. In particular, we are interested in whether it is possible to identify \lq\lq maximal twist" from such properties.

We consider the 
case of a degenerate doublet $\psi$ of up ($u$) and down ($d$) quarks. In the
twisted basis, the doublet $\hat{\psi}$ is given by:
\begin{equation}
\hat{\psi}={\rm e}^{-\frac{i}{2}\omega \gamma_5 \tau_3}\psi , \quad
\hat {\bar{\psi}}=\bar{\psi}{\rm e}^{-\frac{i}{2}\omega \gamma_5 \tau_3}, \quad
\psi=\left(\begin{tabular}{c} u \\ d \end{tabular}\right),
\end{equation}
where $\omega$ is the twist angle and $\tau_3$ is the third Pauli matrix. The bare operators in the twisted and physical basis are therefore related to one other as
\begin{eqnarray}
{\hat V^a_{\mu}}&=&\cos(\omega)V^a_{\mu}-\epsilon^{3ab}\sin(\omega)A^b_{\mu},\quad{\hat V^3_{\mu}}=V^3_{\mu},\\
{\hat A^a_{\mu}}&=&\cos(\omega)A^a_{\mu}-\epsilon^{3ab}\sin(\omega)V^b_{\mu},\quad{\hat A^3_{\mu}}=A^3_{\mu},\\
{\hat S^0}&=&\cos(\omega)S^0-2i\sin(\omega)P^3,\label{first}\\
2{\hat P^3}&= &2\cos(\omega)P^3-i \sin(\omega)S^0,\label{second}\\
{\hat P^a}&=& P^a.
\end{eqnarray}
where $a,b=1,2$ (isospin indices) in these equations and the twisted operators are on the left hand side.
$V$, $A$, $S$, and $P$ represent vector, axial-vector, scalar and pseudoscalar operators, respectively. Some operators are the same in the physical and twisted basis while others are mixtures of physical
operators with opposite parity.

In the following, we will consider the vacuum expectation values (VEVs) of scalar (${\hat S^0}$) and pseudoscalar (${\hat P^3}$) TM loop operators. We define $\tilde{M}^{-1}_{u} \equiv M^{-1}_{u} -(1+2i\mu\kappa\gamma_5)^{-1}$, and call it the \lq\lq hopping parameter subtracted" propagator for the \lq\lq up" quark. ($\mu\rightarrow -\mu$ gives \lq\lq down".) The traces of the subtracted and unsubtracted scalar and pseudoscalar operators are then related as
\begin{eqnarray}
{\rm Tr}[ M^{-1}_u] & = & {\rm Tr}[\tilde{M}^{-1}_u] + { {12} \over {(1+4\kappa^2\mu^2)}},\\
{\rm Tr}[ \gamma_5 M^{-1}_u] & = & {\rm Tr}[\gamma_5 \tilde{M}^{-1}_u] - {{24i\kappa\mu}\over{(1+4\kappa^2\mu^2)}}.
\end{eqnarray}
$\tilde{M}^{-1}_u$ or $\tilde{M}^{-1}_d$ are the quantities actually used in the following. We then have, using the identity $\tilde{M}^{-1}_u=\gamma_5 (\tilde{M}^{-1}_d)^{\dagger} \gamma_5$,
\begin{eqnarray}
<({\hat S^0})_{un}>\equiv {\rm Tr}[ \tilde{M}^{-1}_u+ \tilde{M}^{-1}_d] =2{\rm Re (Tr}[\tilde{M}^{-1}_u]),\\
<({\hat P^3})_{un}>\equiv \frac{1}{2}{\rm Tr}[ \gamma_5\tilde{M}^{-1}_u- \gamma_5\tilde{M}^{-1}_d] =i{\rm Im (Tr}[\gamma_5\tilde{M}^{-1}_u]) ,
\end{eqnarray}
where $<\dots>$ denotes a VEV and \lq\lq un" means unrenormalized, so that one need only calculate the real or imaginary parts of the appropriate up quark operators.

Renormalization constants are expected. For the VEVs in the neutral scalar/pseudoscalar sector, one has~\cite{Frez}
\begin{eqnarray}
<({\hat S^0})_{un}>=\cos(\omega)(<{\bar \psi}\psi>_{un}+2a^{-3}C_s(0))  -2a^{-3}C_s(\omega),\label{unscalar}\\
2<({\hat P^3})_{un}>=-i\sin(\omega)\frac{Z_s}{Z_p}(<{\bar \psi}\psi>_{un}+2a^{-3}C_s(0))  +2ia^{-3}C_p(\omega).\label{unps}
\end{eqnarray}
In the above $a$ is the lattice spacing. In addition, $Z_s$ and $C_s(\omega)$ are the scalar multiplicative and twist-dependent additive renormalization constants, and $Z_p$ and $C_p(\omega)$ are similar quantities for the pseudoscalar sector. According to Eqs.(\ref{unscalar}) and (\ref{unps}) there should be an oscillation in the VEV of the scalar or pseudoscalar as a function of twist angle, $\omega$.

In the following we will make use of the linear correlation coefficient in order to understand the mixings of the operators. Given quark-loop operators $O_1$ and $O_2$, the correlation coefficient, $r$ ($-1<r<1$), has the standard form:
\begin{equation}
r={ {<O_1 O_2> - <O_1><O_2>} \over {\sqrt{<(O_1-<O_1>)^2><(O_2-<O_2>)^2>}  } }.\label{r}
\end{equation}
Physically, this represents the product of disconnected amplitudes.

\section{Search for maximal twist in the scalar/pseudoscalar sector}

\begin{figure}[htbp]
\begin{center}
%\vspace{0.2in}
%\includegraphics[width=0.7\textwidth]{twisting3.pdf}
\includegraphics[width=0.7\textwidth]{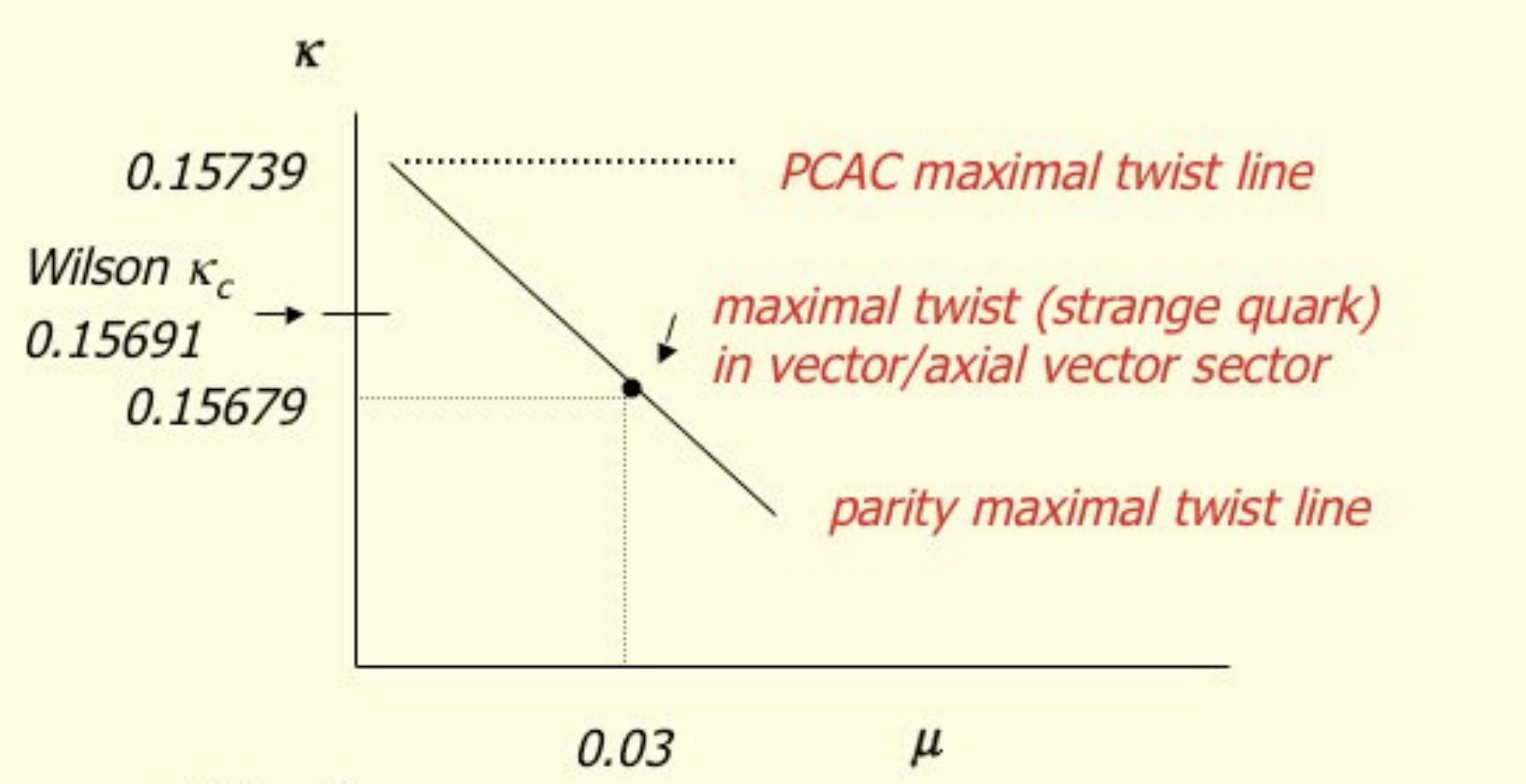}
\caption{Shows our search lines in $\kappa$, $\mu$ space.}
\label{twisting3}
\end{center}
\end{figure}

We conducted a search in parameter space to look for the oscillations in Eqs.(\ref{unscalar}) and (\ref{unps}) in order to identify maximal twist ($\omega=\pm \pi/2$) points in the $\mu$, $\kappa$ plane as well as to understand the mixing patterns of the scalar and pseudoscalar operators. As characterized in Fig.~1 and explained below, we investigated both along $\kappa=0.15739$ and along another line, $\kappa=0.15679$. Our quenched lattices are of size $20^3\times 32$ with $\beta=6.0$.

There are a number of definitions of maximal twist, the most viable being the parity definition~\cite{2} and the PCAC definition~\cite{3}. The parity definition examines the charged vector/axial-vector sector, and finds that maximal twist occurs on a line in the $\kappa$, $\mu$ plane. On the other hand, the PCAC defintion extrapolates the value of $\kappa_c$ obtained from the PCAC relation at non-vanishing bare twisted quark mass $\mu$ to $\mu = 0$. The points at maximal twist defined by this approach lie along a given, fixed $\kappa$ value.

At $\kappa=0.15679$, Ref.~\cite{2} has found that $\mu=0.03$, a point of maximal twist, approximates the strange quark mass. We decided to investigate the twisted scalar and pseudoscalar VEVs along this $\kappa$ value. Note that the connected scalar and pseudoscalar operators are known to mix when evaluated at maximal twist in the connected vector/axial-vector sector~\cite{1}, implying that the maximal twist lines in these two sectors are probably different. 

Our value for the PCAC twist line, $\kappa=0.15739$, was determined from a linear extrapolation to $\mu=0$ of the 4 $\kappa$, $\mu$ pairs given in Ref.~\cite{2} which follow the {\it parity} definition of maximal twist. This value is close to the value $0.157409$ found in Ref.~\cite{jan2} for similar lattices at $\beta=6.0$. Note that the $\kappa_c$ value from the standard Wilson vanishing of the pion mass occurs at $\kappa_c=0.15691$~\cite{Wilson}. Thus, the two investigated values for $\kappa$ bracket the critical Wilson value; see Fig.~1. Therefore, at $\mu=0$ the $\kappa=0.15679$ value is in the standard Wilson phase with small quark mass, whereas the $\kappa=0.15739$ value is in the spontaneous parity-breaking phase known as the Aoki phase~\cite{Aoki}.

\section{Observations and conclusions}

\begin{figure}
\begin{center}
\includegraphics*[viewport=0 0 800 530, scale=0.50]{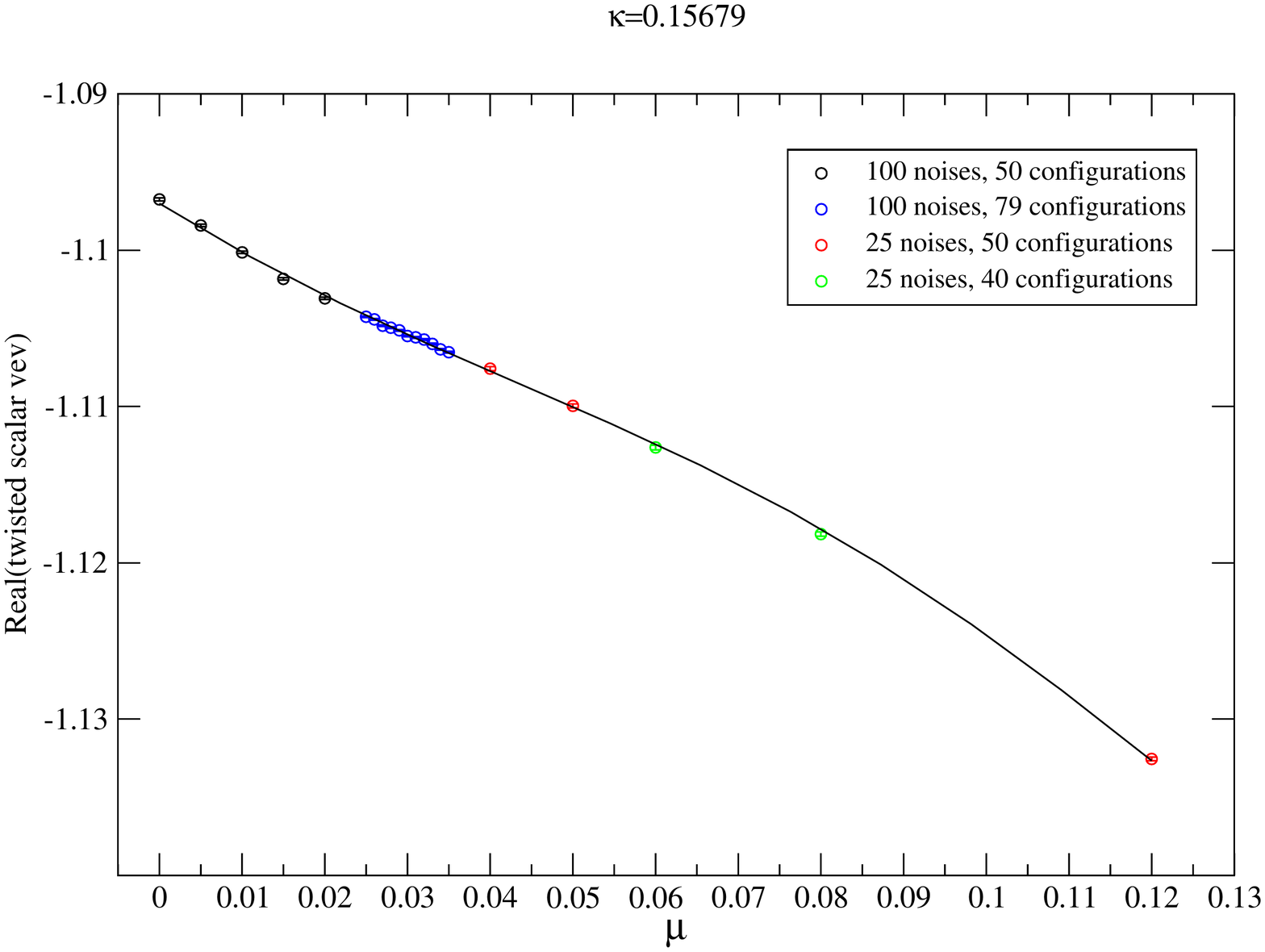}
\caption{A scan across the hopping parameter subtracted VEVs of the twisted scalar operator as a function of the TM parameter, $\mu$, for $\kappa=0.15679$. Fit described in the text.}
\label{scalar-vs-mu-parity}
\end{center}
\end{figure}

Fig.~2 shows the hopping parameter subtracted VEV (per unit volume) of the twisted scalar operator as a function of $\mu$ ($\kappa$ = 0.15679). As pointed out above, there should be an oscillation in the VEV of the scalar or pseudoscalar as a function of twist angle, $\omega$, as one scans at fixed $\kappa$; one is then changing $\omega$ in some non-linear fashion. Indeed, there seems to be a very small oscillation (a {\it twist!}) in $<({\hat S^0})_{un}>$ as a function of $\mu$. However, the inflection point at $\mu \approx 0.044$ is not at the maximal twist point found in the vector, axial-vector case~\cite{2}. The fit of the data
shown is qualitatively useful ($\chi^2$ /(degree of freedom) $\approx 4.5$) but needs improvement. It assumes the twist
angle is directly proportional to $\mu$, which is surely wrong, so the fit can
almost certainly be improved. The fit is given as
\begin{equation}
A+B\mu+C \cos(\frac{\pi \mu}{2D}),
\end{equation}
where $A$, $B$, $C$, $D$ are constants. We found $A=-1.094$, $B=-0.3301$, $C=-0.002845$, and $D=0.04372$. It is this last value that tells us we have an inflection point near $\mu=0.044$.

\begin{figure}
\begin{center}
\includegraphics*[viewport=0 0 800 530, scale=0.50]{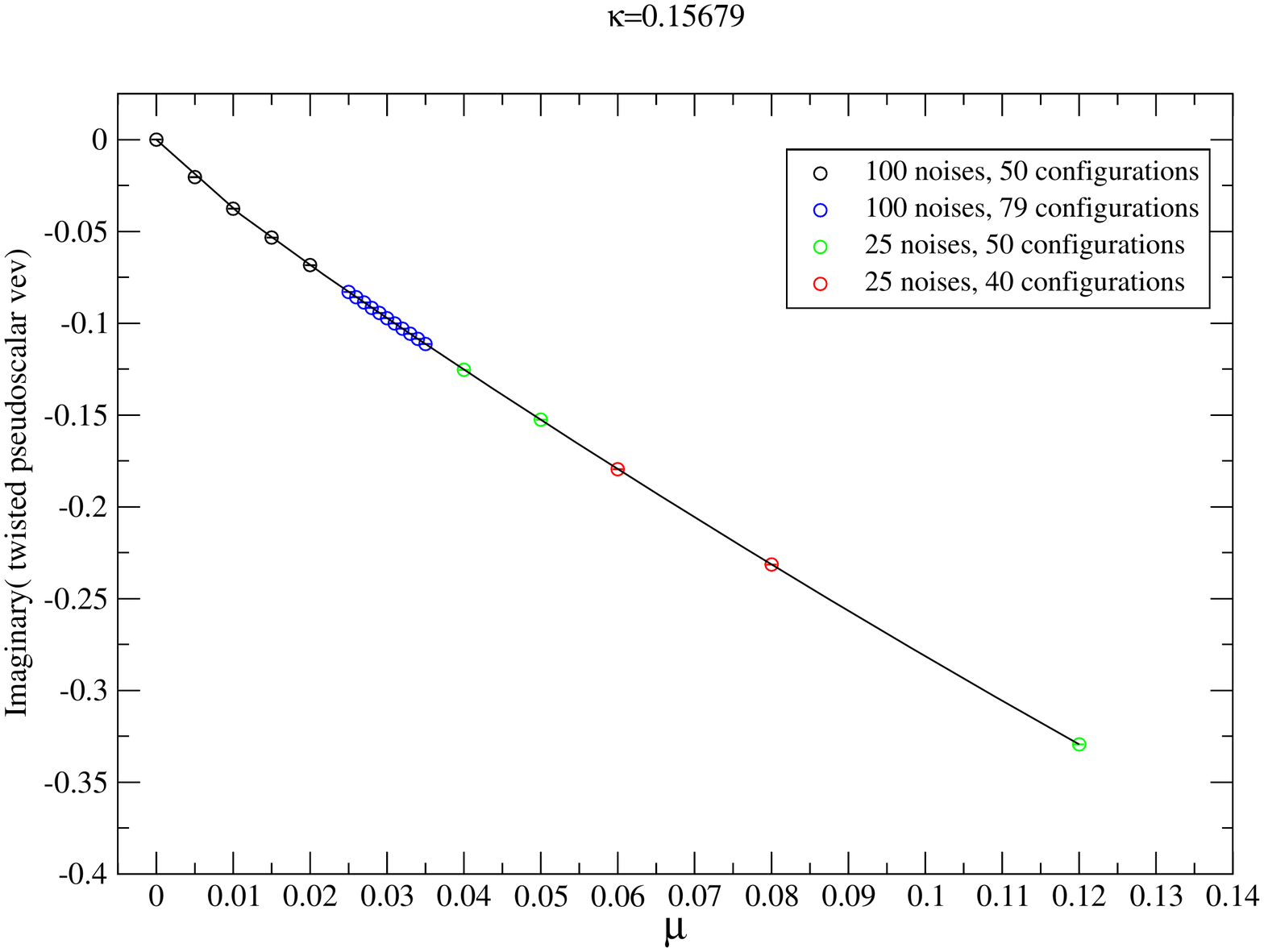}
\caption{A scan across the hopping parameter subtracted VEVs of the twisted pseudoscalar operator as a function of the TM parameter, $\mu$, for $\kappa= 0.15679$. Fit described in the text.}
\label{pscalar-vs-mu-parity}
\end{center}
\end{figure}

Fig.~3 gives the similar VEV of the twisted pseudoscalar operator as a function of $\mu$ ($\kappa=
0.15679$). Unlike the scalar case it is fit well without the need for an oscillatory term; see Eq.(\ref{unps}) for $<({\hat P^3})_{un}>$. The fit we found is actually quite reasonable ($\chi^2$ /(degree of freedom) $\approx 1.7$) and is given by
\begin{equation}
E\mu+F\mu^2+G\tan^{-1}(H\mu),
\end{equation}
where we obtained $E=-2.931$, $F=2.438$, $G=-0.008412$, and $H=164.1$.

Although we do not show the figures here, we also investigated scalar and pseudoscalar operators along the PCAC maximal twist line, $\kappa = 0.15739$. We have much less
data here than in the previous case, and we do not examine the question
of whether oscillatory portions of these functions exist. There is an apparent discontinuity of the scalar and pseudoscalar VEVs at $\mu=0$, and we can no longer fit them with the same functional form used for the smaller $\kappa$ value. As pointed out above, $\mu = 0$
is actually in the Aoki phase.

\begin{figure}
\begin{center}
\includegraphics*[viewport=0 0 800 530, scale=0.50]{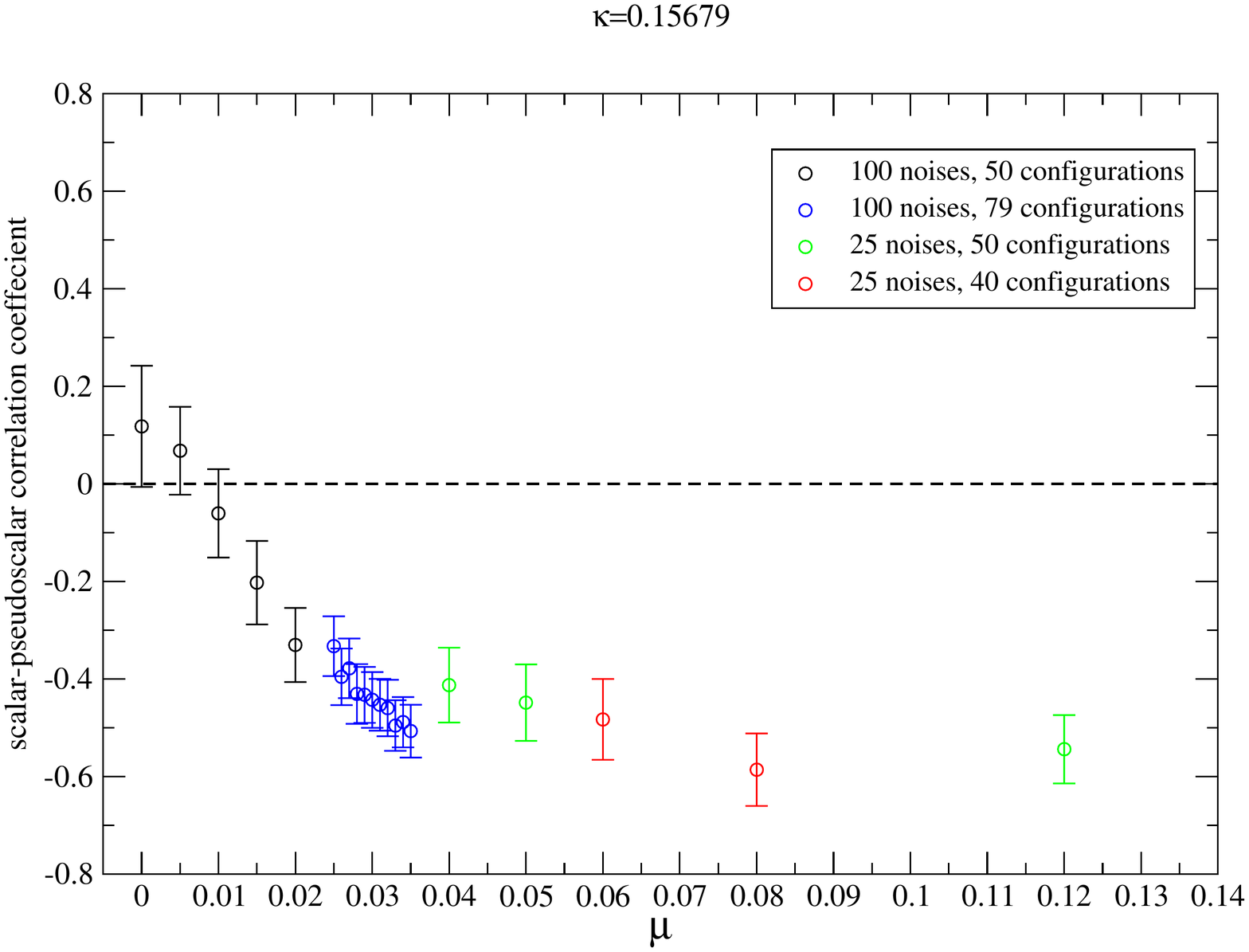}
\caption{Linear correlation coefficient, $r$, of twisted scalar/pseudoscalar operators in a scan across $\mu$ values for $\kappa= 0.15679$.}
\label{correlation-parity}
\end{center}
\end{figure}

\begin{figure}
\begin{center}
\includegraphics*[viewport=0 0 800 530, scale=0.50]{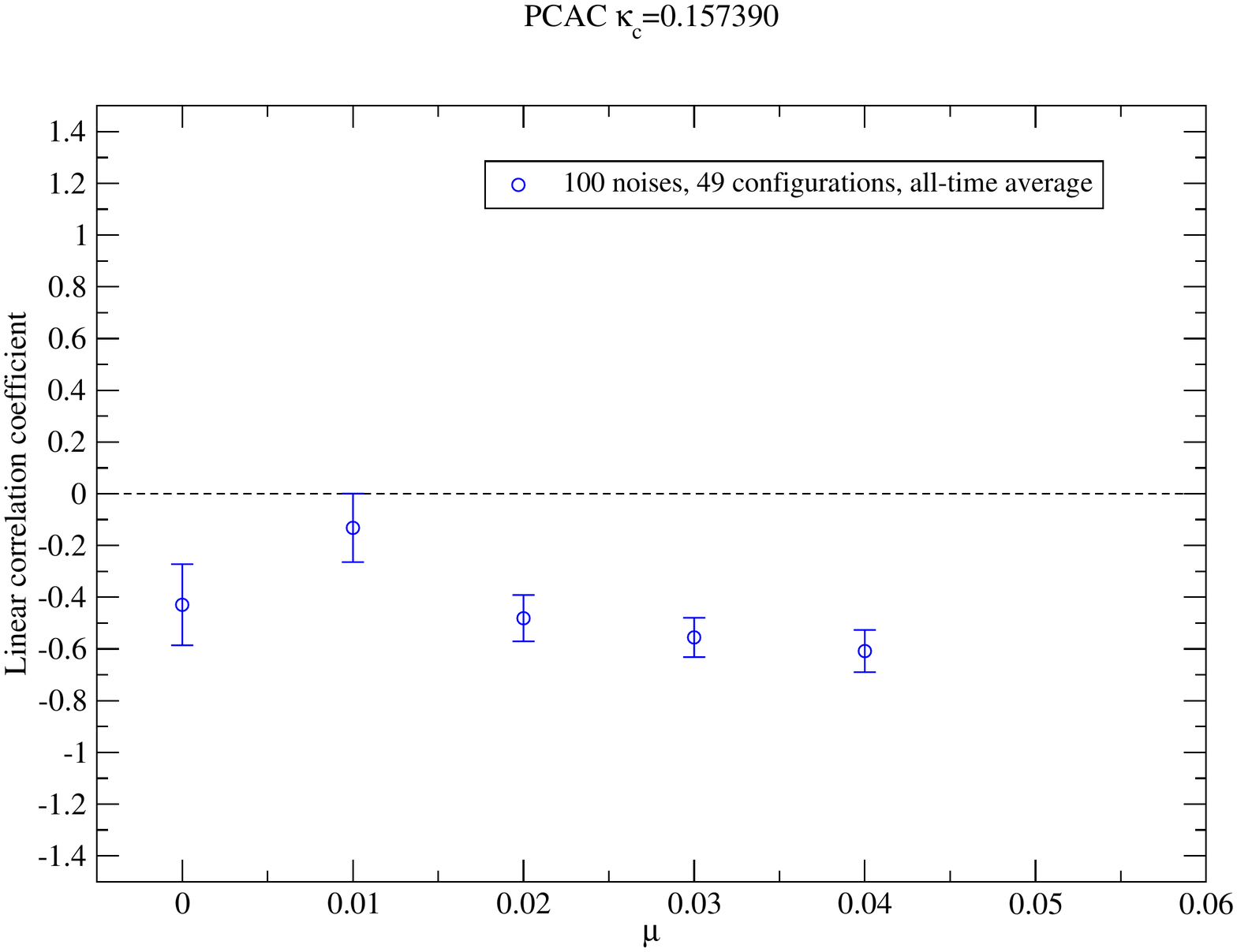}
\caption{Linear correlation coefficient, $r$, of twisted scalar/pseudoscalar operators in a scan across $\mu$ values for $\kappa= 0.15739$ (PCAC case). The $\mu$ = 0 point is in the Aoki phase.}
\label{correlation-pcac}
\end{center}
\end{figure}

To examine the question of the parity of these operators at general $\mu$,
we calculate the linear correlation coefficient, $r$, between twisted scalar and
pseudoscalar operators from Eq.(\ref{r}). Fig.~4 shows the linear correlation coefficient, $r$, for $\kappa= 0.15679$ as a function of $\mu$. This correlation coefficient can be zero only if the parity of the operators is different. The twisted operators are seen to be mixed for all examined values
of $\mu$. The blue symbols in Fig.~4 are values near the parity maximal twist
point at $\mu = 0.03$. The results along the PCAC line, as given in Fig.~5, show a similar pattern.
The point at $\mu=0$ now has a non-zero value for $r$; it is in the Aoki spontaneous parity breaking phase. In Fig.~4 we were able to break the lattice up into two independent statistical parts in order to increase the statistics. It is a \lq\lq half-time average". On the other hand, in Fig.~5 it was necessary to treat the entire lattice volume for each configuration as a single data point in the correlation calculation.
It is an \lq\lq all-time average".

We have also calculated $r$ between {\it Wilson} ($\kappa=0.15679$, $\mu =0$) scalar and pseudoscalar operators and similar twisted operators also at $\kappa = 0.15679$; these 4 graphs are not shown. The Wilson operators have a known parity. The calculation shows that scalar and pseudoscalar TM loop operators have a non-zero overlap {\it only} with Wilson scalar operators.

The conclusion is then that the VEV of the scalar operator, $<({\hat S^0})_{un}>$, seems to be consistent with Eq.(\ref{unscalar}), but that there is no evidence the pseudoscalar VEV, $<({\hat P^3})_{un}>$, contains a $\sin(\omega)$ oscillatory term as in Eq. (\ref{unps}). In addition, both scalar and pseudoscalar TM loop operators, $({\hat S^0})_{un}$ and  $({\hat P^3})_{un}$, are mixed at all $\mu$, and appear to be purely scalar in character.

\section{Acknowledgments}
Calculations were done with HPC systems at
Baylor University and the National Center for Supercomputing Applications. WW thanks the Sabbatical Committee of the College of Arts and Sciences of Baylor University for support.


\begin{thebibliography}{99}
\bibitem{PoS1}{W. Wilcox, \pos{PoS(LATTICE 2007)025}.}
\bibitem{PoS2}{A. Abdel-Rehim, R. Morgan, and W. Wilcox, \pos{PoS(LATTICE 2007)026}.}
\bibitem{Frez}{ALPHA collaboration (R. Frezzotti {\it et al.}), {\em JHEP} {\bf 0108} (2001) 058.}
\bibitem{2}{A. Abdel-Rehim, R. Lewis, and R. Woloshyn, {\em Phys. Rev. D} {\bf 71} (2005) 094505.}
\bibitem{3}{$\chi$LF Collaboration (K. Jansen {\it et al.}), {\em Phys. Lett. B} {\bf 619}  (2005) 184.}
\bibitem{1}{A. Abdel-Rehim, R. Lewis, R. Woloshyn, and J. Wu, {\em Phys. Rev. D} {\bf 74} (2006) 014507.}
\bibitem{jan2}{$\chi$LF Collaboration (K. Jansen {\it et al.}), {\em Phys. Lett. B} {\bf 624}  (2005) 334.}
\bibitem{Wilson}{$\chi$LF Collaboration (K. Jansen {\it et al.}), {\em Phys. Lett. B} {\bf 586}  (2004) 432.}
\bibitem{Aoki}{S. Aoki, {\em Phys. Rev. D} {\bf 30} (1984) 2653.}

\end{thebibliography}
\end{document}